\RequirePackage{tikz}
\RequirePackage{forest}
\documentclass[sn-basic]{sn-jnl}% Math and Physical Sciences Reference Style
\usepackage[bottom]{footmisc}
\usepackage{moreverb,url}
\usepackage{array}

\pgfdeclarelayer{background}
\usepackage{smartdiagram}
\usepackage{bm}
\usepackage{tikz}
\usepackage[center]{caption}
\usetikzlibrary{arrows.meta, shapes.geometric, calc, shadows, positioning,backgrounds,fit}
\usepackage{forest}
\usepackage{stackengine}
\def\shft#1{\stackon[150pt]{}{\kern -0pt #1}}
\def\shfta#1{\stackon[100pt]{}{\kern -0pt #1}}
\def\shfty#1{\stackon[100pt]{}{\kern -0pt #1}}
\def\shftya#1{\stackon[70pt]{}{\kern -0pt #1}}
\def\shftys#1{\stackon[40pt]{}{\kern -0pt #1}}
\def\shftysa#1{\stackon[10pt]{}{\kern -0pt #1}}
\tikzset{
  materia1/.style={draw, fill=fill1, text width=6.5em, text centered, minimum height=1.5em,drop shadow},
  materia2/.style={draw, fill=fill2, text width=6.5em, text centered, minimum height=1.5em,drop shadow},
  materia3/.style={draw, fill=fill3, text width=6.5em, text centered, minimum height=1.5em,drop shadow},
  etape/.style={materia1, text width=6.5em, minimum width=6em, minimum height=3em, rounded corners, drop shadow},
  etape2/.style={materia2, text width=6.5em, minimum width=6em, minimum height=3em, rounded corners, drop shadow},
  etape3/.style={materia3, text width=6.5em, minimum width=6em, minimum height=3em, rounded corners, drop shadow},
  linepart/.style={draw, thick, color=black!50, -LaTeX, dashed},
  line/.style={draw, thick, color=black!90, -LaTeX},
  ur/.style={draw, text centered, minimum height=0.01em},
  back group/.style={fill=yellow!20,rounded corners, draw=black!50, dashed, inner xsep=15pt, inner ysep=10pt},
  set/.style={draw,rectangle,inner sep=0pt,align=left, above=8pt},
}
\tikzset{
  groupassum/.style={draw, fill=ibmlight1, text width=6.5em, text centered, minimum height=3em},
  assum/.style={draw, fill=ibmlight13, text width=6.5em, text centered, minimum height=3em, rounded corners},
  groupcompon/.style={draw, fill=ibmlight5, text width=7em, text centered, minimum height=3em},
  compon/.style={draw, fill=ibmlight52, text width=7em, text centered, minimum height=3em, rounded corners},
  componprop/.style={draw, fill=ibmlight52, text width=9em, text centered, text
    depth  = 1.5cm, minimum height=6em, rounded corners},
  property/.style={draw, fill=ibmlight53, text width=7em, text centered, minimum height=2em},
  grouptask/.style={draw, fill=ibmlight2, text width=7em, text centered, minimum height=3em},  
  task/.style={draw, fill=ibmlight22, text width=7em, text centered, minimum height=3em, rounded corners},
  methods/.style={draw, fill=ibmlight23, text width=7em, text centered, minimum height=3em, rounded corners},  
  ass/.style={assum, text width=6.5em, minimum width=6em, minimum height=3em},
datassum/.style={draw, fill=ibmlight1, text width=7em, text centered, minimum height=3em, rounded corners},
dgpassum/.style={draw, fill=ibmlight12, text width=7em, text centered, minimum height=3em, rounded corners}, 
causalassum/.style={draw, fill=ibmlight14, text width=7em, text centered, minimum height=3em, rounded corners}
}

\usepackage{tikz}
\usetikzlibrary{arrows.meta, shapes.geometric, calc, shadows, positioning,backgrounds,fit}
\usetikzlibrary{arrows}

\tikzset{%
  materia/.style={draw, fill=anniefill1, text width=6.5em, text centered, minimum height=3em,drop shadow, rounded corners},
  materia4/.style={draw, fill=anniefill2, text width=6.5em, text centered, minimum height=3em,drop shadow,  rounded corners},
  materia5/.style={draw, fill=anniefill3, text width=6.5em, text centered, minimum height=1.5em,drop shadow},
  etape/.style={materia, text width=6.5em, minimum width=6em, minimum height=3em, rounded corners, drop shadow},
  linepart/.style={draw, thick, color=black!50, -LaTeX, dashed},
  line/.style={draw, thick, color=black!90, -LaTeX},
  line2/.style={draw, thick, color=black!50, -LaTeX},
  ur/.style={draw, text centered, minimum height=0.01em},
  back group/.style={fill=yellow!20,rounded corners, draw=black!50, dashed, inner xsep=15pt, inner ysep=10pt},
  set/.style={draw,rectangle,inner sep=0pt,align=left, above=8pt},
}

\tikzset{
  task/.style={draw, fill=ibmlight2, text width=5em, text centered, minimum height=2em, rounded corners, , font=\small}
}

\newcommand{\etape}[2]{node (p#1) [materia] {#2}}

\definecolor{anniefill1}{HTML}{CFD5E8}
\definecolor{anniestroke1}{HTML}{324068}
\definecolor{anniefill2}{HTML}{F3F6FB}
\definecolor{anniestroke2}{HTML}{8594B6}
\definecolor{anniefill3}{HTML}{E5E5E5}
\definecolor{anniestroke3}{HTML}{4A4849}
\definecolor{annieline}{HTML}{818689}

\definecolor{ibmlight1}{HTML}{E2EAFD}
\definecolor{ibmlight12}{HTML}{D7E0F9}
\definecolor{ibmlight13}{HTML}{B1C5F9}
\definecolor{ibmlight2}{HTML}{C1B8E8}
\definecolor{ibmlight3}{HTML}{F1CDDF}
\definecolor{ibmlight4}{HTML}{F3DDCF}
\definecolor{ibmlight5}{HTML}{FDEBC3}

\definecolor{ibmlight22}{HTML}{EBE8F7}
\definecolor{python}{HTML}{F7D58A}

\usetikzlibrary{arrows}
\newcommand{\mycomment}[1]{}

\definecolor{ibm1}{HTML}{648FFF}
\definecolor{ibm2}{HTML}{785EF0}
\definecolor{ibm3}{HTML}{DC267F}
\definecolor{ibm4}{HTML}{FE6100}
\definecolor{ibm5}{HTML}{FFB000}

\definecolor{fill1}{HTML}{CFD5E8}
\definecolor{stroke1}{HTML}{324068}
\definecolor{fill2}{HTML}{F3F6FB}
\definecolor{stroke2}{HTML}{8594B6}
\definecolor{fill3}{HTML}{E5E5E5}
\definecolor{stroke3}{HTML}{4A4849}
\definecolor{line}{HTML}{818689}

\definecolor{ibmlight1}{HTML}{F4F7FE}
\definecolor{ibmlight12}{HTML}{E2EAFD}
\definecolor{ibmlight13}{HTML}{D7E0F9}
\definecolor{ibmlight14}{HTML}{C0CFF8}
\definecolor{ibmlight2}{HTML}{F7F5FB}
\definecolor{ibmlight22}{HTML}{EBE8F7}
\definecolor{ibmlight23}{HTML}{e0dcf3}

\definecolor{ibmlight5}{HTML}{FEFBF4}
\definecolor{ibmlight52}{HTML}{fdf3db}
\definecolor{ibmlight53}{HTML}{FCEBC3}

\definecolor{ibmlight3}{HTML}{f7e9f0}
\definecolor{ibmlight4}{HTML}{f7ebe2}
\jyear{2022}%

\theoremstyle{thmstyleone}%
\theoremstyle{thmstyletwo}%

\theoremstyle{thmstylethree}%

\begin{document}

\title[Multi-Agent Influence Diagrams to Hybrid Threat Modeling]{Multi-Agent Influence Diagrams to Hybrid Threat Modeling}

\author[1,2]{\fnm{Maarten} \sur{Vonk}}

\author[2]{\fnm{Anna V.} \sur{Kononova}}

\author[2]{\fnm{Thomas} \sur{Thomas B\"ack}}

\author[1]{\fnm{Tim} \sur{Sweijs}}

\affil[1]{\orgname{The Hague Centre of Strategic Studies}, \orgaddress{\street{Lange Voorhout 1}, \city{The Hague}, \postcode{2514EA}, \country{Netherlands}}}

\affil[2]{\orgdiv{LIACS}, \orgname{Leiden University}, \orgaddress{\street{Einsteinweg 55}, \city{Leiden}, \postcode{2333CC} \country{Netherlands}}}

\abstract{Western governments have adopted an assortment of counter-hybrid threat measures to defend against hostile actions below the conventional military threshold. The impact of these measures is unclear because of the ambiguity of hybrid threats, their cross-domain nature, and uncertainty about how countermeasures shape adversarial behavior. This paper offers a novel approach to clarifying this impact by unifying previously bifurcating hybrid threat modeling methods through a (multi-agent) influence diagram framework. The model balances the costs of countermeasures, their ability to dissuade the adversary from executing hybrid threats, and their potential to mitigate the impact of hybrid threats. We run 1000 semi-synthetic variants of a real-world-inspired scenario simulating the strategic interaction between attacking agent A and defending agent B over a cyber attack on critical infrastructure to explore the effectiveness of a set of five different counter-hybrid threat measures. Counter-hybrid measures range from strengthening resilience and denial of the adversary's ability to execute a hybrid threat to dissuasion through the threat of punishment. Our analysis primarily evaluates the overarching characteristics of counter-hybrid threat measures. This approach allows us to generalize the effectiveness of these measures and examine parameter impact sensitivity. In addition, we discuss policy relevance and outline future research avenues. }

\keywords{Hybrid Threats, Adversarial Behaviour Analysis, Strategic Interaction Modeling, Deterrence, Influence Diagrams, Causal Inference, Game Theory}

\maketitle

% Anna: Maarten, please consider all my comments purely as suggestions (even more so if I include a question mark in the text of the comment)
\section{Introduction}
Hybrid threats, defined as the coordinated use of violent and non-violent means to exploit vulnerabilities and influence adversaries below the threshold of armed conflict, pose an escalating challenge in an era of growing global interconnectedness. In response, states have implemented a broad spectrum of potential counter-hybrid measures, including economic sanctions, cyber defense strategies, information campaigns, and diplomatic initiatives. However, the effectiveness of these measures remains uncertain due to the complex and opaque nature of hybrid threats, which often operate across multiple domains and resist attribution. Therefore, researchers have resorted to modeling approaches, employing either game theoretic \citep{balaban2018hybrid} or probabilistic methods \citep{balcaen2022game}, as they offer a means to address this uncertainty by systematically evaluating the impact of different measures under varying conditions while providing actionable insights into the interplay between threats and countermeasures.

Building on these methodologically distinct approaches, this paper proposes an an integrated probabilistic and game-theoretic model to comprehensively assess the effectiveness of counter-hybrid threat measures. The interaction between two state-like agents is modeled probabilistically to account for cognitive and psychological deterrence factors \citep{berejikian2002cognitive}, and game-theoretically to capture strategic decision-making. The defender’s pay-off is based on the balance between the costs of countermeasures and the potential damage from hybrid attacks, accounting for the possibility that such attacks may or may not have been successfully deterred. Optimal countermeasures are derived by maximizing expected pay-offs while game equilibria are distilled by considering the adversary's strategic responses.

To test the modeling approach, a cyber threat scenario on critical infrastructure was developed, inspired by real-world incidents. Policy experts and available literature were consulted to identify relevant countermeasures to this cyber threat and collate estimates of the cost, damage mitigation ability, and deterrence ability of each of the counter-hybrid measures. These estimates provided a basis for analyzing counter-hybrid measures and allowed us to gauge their effectiveness across different scenarios, including scenarios where the adversary engages in strategic competition. To validate our proposed approach, we contextualized our findings within the framework of existing studies and conducted sensitivity analyses to identify and quantify the most influential variables driving the model's outcomes. This enabled us to address the following research questions:
\begin{itemize}
    \item Which characteristics of counter-hybrid threat measures most effectively enable the defending agent to address the cyber threat on critical infrastructure posed by the attacking agent given uncertainty about 1) the measures' ability to dissuade the adversary from carrying out the attack, 2) the measures' ability to mitigate the impact of the attack, and 3) the measures' cost. 
    \item Which characteristics of counter-hybrid threat measures within the context of a cyber threat on critical infrastructure can cause the strategies of players to form an equilibrium given 1) the costs of counter-hybrid measures and hybrid operations, 2) the measures' ability to mitigate the impact of the attack, and 3) a strategically operating adversary. 
\end{itemize}

In answering these research questions, this article is structured as follows: the following section introduces background information on deterrence and hybrid threats. Thereafter, we outline the modeling methodology along with a discussion of the transformation of insights from the literature review and expert opinion into input for the model via probability distributions. The subsequent section describes the content of the hybrid threat scenario and the associated cross-domain counter-hybrid measures. The results are then presented and analysed in the penultimate section, where they are contextualised within a broader framework of similar studies. The final section reflects on the policy relevance of the findings and identifies future research avenues. 

\section{Background} \label{background}
Fast-paced technological developments, deeper economic integration, and the digitally wiring of societies have reshaped contemporary interstate competition, furnishing revisionist states with innovative tools to pursue strategic objectives below the threshold of large-scale armed conflict. In Europe, observers use the term \emph{hybrid threats} to broadly define behaviors corresponding to the ‘‘coordinated and synchronized’’ use of violent and non-violent means, often difficult to detect and attribute, aimed at weaponizing democratic processes and exerting influence over adversaries.\footnote{Hybrid CoE, Hybrid Threats as a Concept, Accessed 7 June 2023,  \url{https://www.hybridcoe.fi/hybrid-threats-as-a-phenomenon/}} Although often used interchangeably, the term ‘hybrid conflict’ is conceptually different from what U.S. scholars and strategists refer to as ‘grey zone’ strategy \citep{centroconfessions}. Grey zone strategies refer to a peculiar condition of quasi-persistent interstate conflict in which aggressive operations apparently covered by legal justifications are used to coerce adversaries and pursue limited objectives, exploiting adversaries' vulnerabilities below the threshold of detection and attribution \citep{morris2019gaining}. Both terms however refer to aggressive behavior below the threshold of conflict that includes information and psychological operations, political, diplomatic and economic coercion, offensive cyber operations, and the use of proxies to destabilize adversaries.

States currently seek to devise a range of counter-hybrid policies aimed at increasing resilience and imposing costs on rival states in order to deter such behaviors. However, when compared to openly aggressive or offensive conducts in the conventional and nuclear domains, hybrid operations pose unique challenges that, because of their opaque and cross-domain nature, prove difficult to deter and defend against \citep{mallory2018new}.

Traditionally, deterrence is about ‘‘discouraging or restraining a nation-state from taking unwanted actions’’ \citep{Mazarr2021} by either denying an adversary the ability to achieve its objectives (deterrence by denial) \citep{wilner2021deterrence,  viad036, arce2023cybersecurity} or threatening to impose cost following the actions (deterrence by punishment) \citep{FreedmanLawrenceDavid2004D, george1974deterrence}. In writings on conventional and nuclear deterrence, it is widely acknowledged that perceptions are key to successful deterrent efforts \citep{Mazarr2021} in the sense that the adversary must perceive ‘‘that the costs likely to be incurred from his initiative will outweigh the potential gains’’ \citep{nyemann2018going}. To achieve their objective, deterrent efforts should be clear, proportional, and credible. Clarity entails the ability to communicate unambiguously which measures the defender will likely adopt to respond; proportionality describes the equivalence between the means used to deter and the objectives pursued by the defender; finally, credibility - which is a function of clarity and proportionality - is rooted in the deterrer's capability and willingness to act in face of external aggression \citep{Mazarr2021}. Therefore, traditional tenets of deterrence rest on a credible threat of punishment or the credible denial of gains.

The application of these favoring conditions of classical deterrence is significantly hampered in the hybrid context. First, aggressive behaviors in the grey zone do not simply materialize as military confrontation but rather as a complex combination of military and non-military, overt and covert, operations involving economic coercion, disinformation campaigns, offensive cyber operations, and even the deployment of armed groups.\footnote{NATO’s response to hybrid threats, Accessed 19 April 2023, \url{https://www.nato.int/cps/en/natohq/topics\_156338.htm}} Second, hybrid threats present additional and unique problems that significantly blur the way in which confrontations take place. Among many, two of the most pressing issues in tackling hybrid threats derive from the fact that grey zone activities occur on a continuous basis and are often difficult to attribute to a specific adversary. This makes deterrence even more complicated \citep{sweijs2021essence}. 

Traditional in-domain punishment and denial strategies, while maintaining some relevance, are no longer sufficient to address the complex interactions occurring in the grey zone \citep{10.1093/oso/9780190908645.001.0001}. Traditional deterrence strategies should evolve into complex cross-domain strategies that - in addition to the threat of costs and the denial of potential gains - include the provision of reassurances and incentives to the adversary (assurance), the promotion of international cooperation and norm development (norms), and the exploitation of economic and systemic (inter)dependencies (entanglement) to influence adversarial behaviors. In addition, given the perpetual state of tension in the grey zone, deterrent efforts should be ‘cumulative’: defenders should consider counter-hybrid strategies as a ‘‘longer-term process
in which a one-off transgression does not spell failure’’ and in which adversarial behavior ‘‘is
shaped by the deterrer in a concerted effort’’\citep{sweijs2019cross, tor2017cumulative}. In this sense, the authors suggest that a broader strategy of dissuasion is better equipped to address hybrid threats. Dissuasion is hereby understood as the overarching strategy encompassing both punishment and denial responses with advanced countermeasures that can leverage political, diplomatic, and economic relations among peers \citep{nye2016deterrence}. Consequently, to dissuade adversaries in the grey zone, counter-hybrid strategies can only be effective if all instruments of state power across the Diplomatic-Information-Military-Economic-Financial-Intelligence-Law enforcement (DIMEFIL) spectrum are strategically deployed, while simultaneously managing escalation dynamics and potential retaliatory responses from adversaries.\footnote{Strategic Communications Hybrid Threats Toolkit, Monika Gill, Ben Heap, and Pia Hansen (NATO Strategic Communications
Centre of Excellence), 30–36, accessed 26 April 2023, \url{https://stratcomcoe.org/publications/strategic-communications-hybrid-threats-toolkit/213.
}} 

In practice, however, there are no agreed-upon principles, metrics or guidelines to craft successful cross-domain responses and the risk of disproportional and ambiguous behaviors is ever-present \citep{sweijs2021essence}. Since deterrence and dissuasion are ultimately rooted in perception, the effectiveness of counter-hybrid strategies is dependent on real-world motivations and perceived core interests of the adversary. This includes its propensity for offensive action as well as the vulnerabilities it seeks to protect from possible retaliation.\footnote{The Nine Commandments on Countering Hybrid Threats Threats, Internationale Politik Quarterly, Michael Rühle, accessed 20 May 2023, \url{https://ip-quarterly.com/en/nine-commandments-countering-hybrid-threats}} In most situations, however, policymakers lack sufficient information regarding the broader strategic objectives that adversaries pursue in the grey zone; what decision-making processes and pay-off calculus drive operations below the threshold of war; and, to what extent counter-hybrid policies affect hybrid threat behavior, not in the least, because a vast array of hybrid threats takes place below the threshold of detection and attribution \citep{kilcullen2019evolution}. Hence, it is extremely difficult to gauge the effectiveness of counter-hybrid policies in the real world.

%Will add literature review here!
While attempting to model hybrid threat dynamics, some authors have resorted to game theory to examine strategic interactions among rival states in an effort to overcome the paucity of information available  \citep{balcaen2022game, attiah2018game}. Others have incorporated scarce data sources into Bayesian modeling techniques \citep{davis2021influencing, balaban2018hybrid} with the aim of refining domain knowledge with available data. Although current game-theoretic approaches struggle to capture the complexities and uncertainties inherent in hybrid threat dynamics, Bayesian modeling techniques, while effective at handling uncertainty, fall short in representing strategic interactions. 

This paper introduces a novel contribution by proposing an influence diagram approach that models the deep uncertainties of counter-hybrid policies (e.g., threat detection, attribution, and mitigation effects) as probabilistic relations \citep{howard2005influence}. By extending this approach to a multi-agent framework \citep{KOLLER2003181}, it integrates game-theoretic considerations, offering a unified methodology that not only bridges these two approaches but also enhances their applicability to complex, multi-actor hybrid threat scenarios.

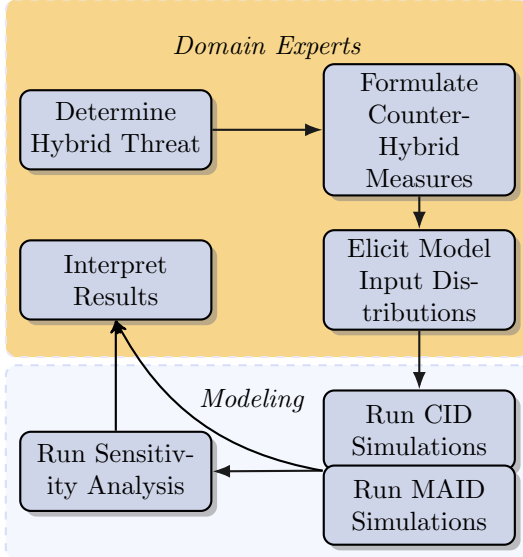
\begin{figure}[!t]
\centering
\begin{tikzpicture}[->,>=stealth',auto,node distance=3cm,
  thick,main node/.style={circle,draw,font=\sffamily\Large\bfseries}]

  % Draw diagram elements
  \path \etape{1}{Determine Hybrid Threat};
  \path (p1)+(4,0) \etape{2}{Formulate Counter-Hybrid Measures};
  \path (p1)+(4,-2) \etape{3}{Elicit Model Input Distributions};
  \path (p1)+(4,-4) \etape{4}{Run CID Simulations};
  \path (p1)+(4,-5) \etape{8}{Run MAID Simulations};
  \path (p1)+(0, -4.55) \etape{5}{Run Sensitivity Analysis};
  \path (p1)+(0, -2) \etape{6}{Interpret Results};
  
  \path (p1)+(2,1.1) node (python){\textit{Domain Experts}};
  \path (p1)+(1.8,-3.6) node (embed) {\textit{Modeling}};

  % Draw arrows between elements
  \path [line] (p1.east) -- coordinate[pos=0.5] (a1) node [above]  {} (p2.west);
  \path [line] (p2.south) -- coordinate[pos=0.5] (a1) node [above]  {} (p3.north);
  \path [line] (p3.south) -- coordinate[pos=0.5] (a1) node [above]  {} (p4.north);
  \path [line] (p4.south west) -- coordinate[pos=0.5] (a1) node [above]  {} (p5.east);
  \path [every node/.style] (p5.north) edge node [above] {} (p6.south);
  \path [every node/.style] (p4.south west) edge[bend left=26] node [above] {} (p6.south);
    
  \begin{scope}[on background layer]
    \node (bk2) [back group] [fit=(p1) (p2) (python) (p3) (p6), inner xsep=5pt, inner ysep=10pt, draw, thick, ibmlight22, fill=python] {};
    \node (bk3) [back group] [fit=(p4) (p5) (p8) (embed), inner xsep=5pt, inner ysep=5pt, draw, thick, ibmlight13, fill=ibmlight1] {};
  \end{scope}
\end{tikzpicture}
\caption{The figure illustrates the processes involved in counter-hybrid threat analysis as proposed in this study. Initially, domain experts identify the hybrid threat and develop corresponding counter-hybrid measures. They also provide key input parameters, which are used to construct probabilistic input distributions for the model. Samples from these distributions are used to run simulations with the causal influence diagram (CID) as well as the multi-agent influence diagram (MAID) model. Finally, a sensitivity analysis is performed and the model results are interpreted and compared with existing studies.}
\label{scope_paper}
\end{figure}

\section{Methodology} \label{sectmeth}

This section outlines the underlying mechanism of the simulation model, along with the process of eliciting inputs required to run simulations using the model. The full scope of the proposed method, including the extent of expert involvement, is illustrated in Figure \ref{scope_paper}.

The model considers the behavior of two agents possessing the characteristics of sovereign states, following the two-agent approach of Balcaen et al.~\cite{balcaen2022game} and Attiah et al.~\cite{attiah2018game}. On one side, agent A aims to pursue its strategic objectives using hybrid attacks. On the other side, agent B wishes to protect its national interests and deter and defend against hybrid attacks. Agent B, the defender, chooses a counter-hybrid posture to deter or dissuade agent A from carrying out a hybrid operation. For this reason, the strategy is referred to as a counter-hybrid measure. To this purpose, agent B explores available counter-hybrid measures to dissuade the adversary from carrying out hybrid attacks by altering the cost-benefit calculus \citep{8405009}. The defender may also adopt measures - such as the enhancement of detection and/or attribution capabilities \citep{schneider2019deterrence} - that would boost resilience and mitigate the potential impact of hybrid conducts \citep{GiladAmitai2023MtRo}. 

Both the counter-hybrid measure and the hybrid attack bear direct costs. Such costs represent not only the resource costs but also costs involving for instance political capital to rally domestic and international support as well as potential costs associated with escalation. The interaction between agents A and B is of a zero-sum nature. Probabilities are used to reflect the considerable degree of uncertainty over the value of key variables that lead to different outcomes. Examples are uncertainty associated with the impact of counter-hybrid measures on the strategic calculus of agent A, as well as with detection, attribution, and the mitigatory impact of the counter-hybrid measure.

First, a causal influence diagram approach is introduced, enabling the optimization of counter-hybrid deterrence strategies when the adversary's responsiveness is estimated probabilistically. This approach is then extended into a multi-agent influence diagram by modeling both agents as players within a game-theoretical framework, allowing for the analysis of game equilibria.

\subsection{Optimizing Counter-Hybrid Strategy}
A Bayesian Network modeling technique is often used to account for the combination of probabilistic and deterministic relationships \citep{balaban2018hybrid, zhu2020dual, yan2012towards, johansson2008bayesian, wang2012air}. This allows for the factorization of the joint distribution as the product of the conditional probabilities according to the structure of the graph \citep{Pearl2009}. Bayesian networks can be extended to causal influence diagrams to further dissect the nodes of the graph into random variables, utility nodes, and decision nodes \citep{howard2005influence} and allow for causal interventions. In the first proposed modeling approach, the defender preemptively commits to a selected counter-hybrid posture to deter or dissuade agent A from conducting a hybrid operation, represented by a decision node. The adversary's response is driven by the estimated probability of successful dissuasion.

\begin{figure}[!t]
\centering
\begin{tikzpicture}

  % Draw diagram elements
  \path \etape{1}{Counter-Hybrid Measure $D$};
  
  \path (p1)+(3,0) \etape{2}{Hybrid Conduct $C$};
  \path (p1)+(1.5,-1.5) \etape{3}{Pay-off $\Theta$};

  \path (p1)+(-0.75,-3.5) \etape{4}{Counter-Hybrid Costs $\Gamma$};
  \path (p1)+(3.75,-3.5) \etape{5}{Hybrid Conduct Costs $H$};

  \path (p1)+(0,-5.5) \etape{6}{Agent B Total Pay-off $DP$};
  \path (p1)+(3,-5.5) \etape{7}{Agent A Total Pay-off $CP$};
  
  % Draw arrows between elements
 % \path [line] (p1.south) -- coordinate[pos=0.5] (a1) node [above]  {} (p1);
  \path [line] (p1.south) -- node [above] {}  (p3.west);
  \path [line] (p1.east) -- node [above] {}  (p2.west);
  \path [line] (p2.south) -- node [above] {}  (p3.east);
  
  \path [line2] (p1.south) -- node [above] {}  (p4.north);
  \path [line2] (p2.south) -- node [above] {}  (p5.north);
  \path [line2] (p3.south) -- node [above] {}  (p6.north east);
  \path [line2] (p3.south) -- node [above] {}  (p7.north west);
  \path [line2] (p5.south) -- node [above] {}  (p7.north);
  \path [line2] (p4.south) -- node [above] {}  (p6.north);  

  \begin{scope}[on background layer]
    \node (bk1) [back group] [fit=(p1) (p2) (p3), anniefill3, fill=anniestroke2] {};
    \node (bk2) [back group] [fit=(p4) (p5) (p6) (p7), inner xsep=15pt, inner ysep=10pt, draw, thick, anniefill1, fill=anniestroke1, rounded corners] {};
  \end{scope}
\end{tikzpicture}
\caption{(Multi-Agent) Causal Influence Diagram encoding hybrid threat modeling. While the bottom background layer groups the deterministic variables, the top layer represents the probabilistic variables. Probabilistic relations are displayed by black arrows and deterministic relations by grey arrows.}
\label{Figure1}
\end{figure}
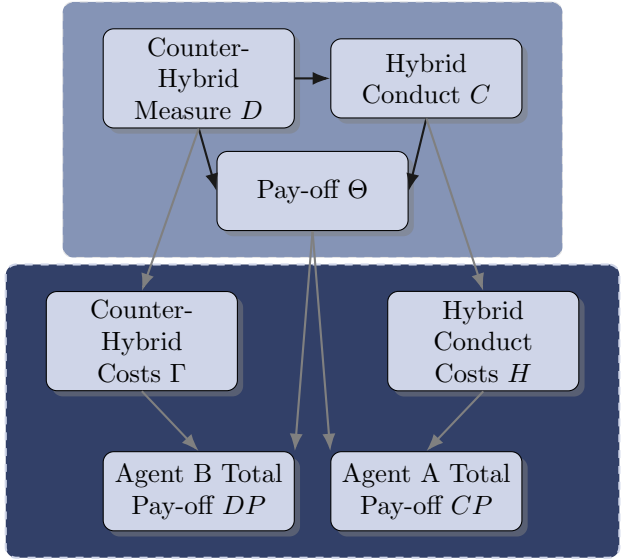

More formally, let $D$ be the decision variable describing the counter-hybrid measure, which is followed by $C$, the random variable for the hybrid conduct. The available strategies and state spaces  for $D$ and $C$ are $\Omega_D = \{d_1, \dots,d_n\}$ and $\Omega_C= \{c_1, \dots,c_m\}$ respectively. Each combination of counter-hybrid measure and hybrid operation variables leads to a discrete pay-off random variable $\Theta$ with state space $\Omega_\theta = \{\theta_1, \dots,\theta_k\}$ representing the interaction pay-off structure excluding direct costs. Because we separate the pay-off structure from hybrid and counter-hybrid interaction from the direct cost of hybrid operation and counter-hybrid measure, we assume the pay-off structure has a zero-sum game component, meaning a positive value $\theta \in \Theta$ corresponds to the gain of agent B and the loss of agent A, while a negative value for $\theta \in \Theta$ represents a gain for agent A and the loss for agent B. The direct costs for the counter-hybrid measure and the hybrid attack are drawn from a cost probability function and denoted by $\Gamma$ and $H$ respectively, where the costs for the counter-hybrid measure has state space $\Omega_\Gamma=\{\gamma_1,\dots, \gamma_n\}$ with $\gamma_i\geq 0$ for $i=1,\dots,n$ and the costs for the hybrid attack has state space $\Omega_H=\{\eta_1,\dots, \eta_m\}$ with $\eta_i\geq 0$ for $i=1,\dots,m$. The total pay-off can finally be calculated by $DP = \Theta-\Gamma$ for agent B and $CP = -\Theta-H$ for agent A. The causal influence diagram that corresponds with these relations is depicted in Figure \ref{Figure1}.\footnote{While the cost variables are priorly drawn from probability distributions, they become deterministic in the influence diagram, as only one cost value corresponds to a (counter-) hybrid measure per experiment.} Note that factorizing the direct costs in the initial pay-off node could have made the influence diagram purely probabilistic. However, for clarification purposes, direct costs have been separated from interaction costs. All probability distributions in the influence diagram are typically assumed to be categorical, which makes the conditional probabilities expressable via conditional probability tables (CPTs). According to the influence diagrams structure, the joint probability can be factorized via:
\begin{align*}
P(d, c, \theta, \gamma, \eta, dp, cp) &= P(d)P(c\mid d)P(\theta \mid d, c)P(\gamma\mid d) \\&P(\eta \mid c)P(dp\mid \gamma, \theta)P(cp\mid \theta, \eta).
\end{align*}

In order to calculate the probability of the total pay-off for agent B, we have to marginalize out all other variables (we can ignore the deterministic variables in other parts of the network \citep{koller2009probabilistic}):

\begin{align*}
P(dp)&=\sum_{d}\sum_{c}\sum_{\theta} \sum_{\gamma}   P(d, c, \theta, \gamma, dp) \\ &=\sum_{d}\sum_{c}\sum_{\theta} \sum_{\gamma} P(d)P(c\mid d)P(\theta \mid d, c) 
\\ & \  P(\gamma\mid d)P(dp\mid \gamma, \theta).
\end{align*}

Suppose that agent B has access to the potential costs of counter-hybrid measures $\Omega_\Gamma=\{\gamma_1,\dots, \gamma_n\}$, their deterrence capacity $P(c\mid d)$ and their ability to mitigate potential damages $P(\theta \mid d, c)$. Agent B's objective is to compute the counter-hybrid measure that maximizes its total pay-off under the assumed probability distributions \citep{Andrew2022}. Specifically, we want to find the intervention $do(D=d)$ that maximizes agent B's expected total pay-off. Since the counter-hybrid measure node has no incoming arrows (see Figure \ref{Figure1}), intervening on a variable is the same as conditioning on this variable \citep{Pearl2009}:

\begin{align*}
&\max_{d_1,\dots,d_n}\mathbb{E}[dp\mid do(D=d)]=\\&\max_{d_1,\dots,d_n}\sum_{c}\sum_{\theta} \sum_{\gamma}dp   P(c, \theta, \gamma, dp, \mid do(D=d))= \\& \max_{d_1,\dots,d_n} \sum_{c}\sum_{\theta} \sum_{\gamma} dp P(c\mid d)P(\theta \mid d, c)
\\&P(\gamma\mid d)P(dp\mid \gamma, \theta).
\end{align*}
 This can be formulated as an integer linear program (ILP):

\begin{equation*}
\begin{array}{rrclcl}
& \displaystyle  \max \sum_{h=1}^k\sum_{j=1}^m\sum_{i=1}^n & \multicolumn{3}{l}{\theta_h w_{ijh} q_{ij} p_{i}+ \displaystyle \sum_{i=1}^n \gamma_{i}p_i} & \\
\textrm{subject to} & \displaystyle \sum_{i=1}^n p_i & = & 1 & & \\
& p_i  & \in & \{0,1\} &  i =1,\dots,n  &  \\
\end{array}
\end{equation*}

\subsection{Subgame Perfect Equilibrium}

\begin{table*}[!t]
\caption{Values drawn from probability distributions. Costs of counter-hybrid measures and damaging impacts of hybrid attacks are expressed in US million dollars. While the costs of counter-hybrid measures and damaging impacts of hybrid attacks are drawn from variants of the normal distribution, the probability values for the ability to deter and the ability to mitigate damaging impacts are drawn from their corresponding conjugate priors (Beta and Dirichlet, respectively). }
\begin{tabular}{|m{2em}|m{12em}| m{14em}|}
\hline
\multicolumn{1}{|l|}{Value} & \multicolumn{1}{l|}{Meaning} & \multicolumn{1}{l|}{Probability Distribution}  \\ \hline \hline
$\theta_h$  & Potential damage of a hybrid attack  & Drawn from different truncated normal distributions for each category of damaging impacts~\citep{norgaard1980expected}. \\ \hline
$\gamma_i$  & Cost for conducting counter-hybrid measure $d_i$  & Drawn from different truncated normal distributions for each counter-hybrid measure $d_i$~\citep{norgaard1980expected}    \\ \hline
$q_{ij}$  & Probability of the adversary conducting hybrid operation $c_j$ after counter-hybrid measure $d_i$  & Drawn from different Beta distributions for each counter-hybrid measure. \\ \hline
$w_{ijh}$  & Probability of potential damage $\theta_h$ after the adversary conducts hybrid conduct $c_j$ and defender counter-hybrid measure $d_i$  & Drawn from different Dirichlet distributions for each counter-hybrid measure and hybrid operation combination~\citep{spiegelhalter1990sequential}.  \\ \hline
\end{tabular}
\label{Table1}
\end{table*}

In optimizing the counter-hybrid strategy, probabilities are used to estimate the likelihood of successfully deterring the adversary after each measure. These probabilities are determined ex-ante, meaning they are drawn before a counter-hybrid measure is chosen. As a result, they do not account for any short-term pay-off adjustments that occur during the interaction that eventually set the outcome in equilibrium. To this end, we extend the causal influence diagram approach to multi-agent influence diagrams (MAIDS) \citep{KOLLER2003181,hammond2021equilibrium} to account for these strategic considerations and address the notion of equilibrium. 

Formally, the hybrid conduct node $C$ of Figure \ref{Figure1} is no longer modeled probabilistically but is instead represented as a decision node. Additionally, the distinction between the two agents, $A$ and $B$, is made explicit by assigning decision and utility nodes to each respective player. Specifically, decision node $D$ and utility node $DP$ are assigned to agent $B$, while decision node $C$ and utility node $CP$ are associated with agent $A$.

We seek a solution concept that pinpoints a subset of possible outcomes when agents act rationally. While the Nash equilibrium (NE) is a widely used solution concept in non-cooperative games, where no agent can gain by changing their strategy unilaterally ~\citep{nash1950equilibrium}, it may lead to non-credible threats. These are decisions made by an agent that would not be in their best interest to execute if the situation were to arise ~\citep{selten1988reexamination}. In the context of the hybrid threat game, non-credible threat equilibria emerge when the attacker threatens to conduct a hybrid operation despite it not being in their best interest in terms of pay-off.

To address this, we adopt the concept of \emph{subgame perfect equilibrium} (SPE) \citep{hammond2021equilibrium}. A subgame perfect equilibrium is defined as a refinement of Nash equilibrium in which the Nash equilibrium conditions are satisfied not only for the overall game but also for every subgame within it. Therefore, the SPE eliminates the existence of non-credible threats and ensures that decisions regarding counter-hybrid measures or hybrid operations remain rational and optimal at every stage of the interaction. Subgame perfect equilibria can be determined by applying backward induction across all identified subgames.

\begin{table*}[!t]
\caption{Five Counter-Hybrid Measures}
\begin{tabular}{|m{5em}|m{5em}| m{9em} | m{10em}|}
\hline

\multicolumn{1}{|l|}{Domain} & \multicolumn{1}{l|}{Title} & \multicolumn{1}{l|}{Capability} & \multicolumn{1}{l|}{Rationale} \\ \hline \hline
Cyber  & Boost cyber resilience at the wider societal level  & Introduce legislation or collaboration that requires individuals and companies to adopt sufficient levels of cyber resilience, based on the specific risk exposure of the subject.  & Boosting cyber resilience at the broader societal level is one of the core tenets of a whole-of-society approach to hybrid/cyber threats. \\ \hline
Cyber & Employ offensive cyber capabilities  & Use offensive cyber operations in order to undermine the target.   & It is often the case that cyber attacks are not directly attributable. In the short term, it provides a covert way to influence the target. The effect of an offensive cyber attack is scalable.  \\ \hline
Legal & Market restrictions & Introduce legislation to restrict an opponent from accessing your market in a specific sector (such as ICT).  & Such measure reduces the possibilities for an adversary to exploit vulnerabilities but it may also clash with other legal commitments (see international trade commitments against market restrictions based on nationality). \\ \hline
Diplomatic & Open deterrence messaging through strategic communications  & Communicate one's strategic posture in order to convince a target to comply with one's strategic aims.  & Being transparent with the hostile actor regarding one's own strategic strengths and possible actions. This increases the possibility of a better-informed decision by the hostile actor.  \\ \hline
\end{tabular}
\label{Table2}
\end{table*}

\subsection{Probability Distributions \& Elicitation}
Filling the influence diagram with accurate conditional probabilities is widely recognized as a challenging task \citep{johansson2008bayesian} and a rigorous elicitation process should be developed to ensure the highest degree of accuracy in the inputs. To maintain a realistic perspective in our estimates, we have attempted to estimate the cost, potential deterrence capacity, (either denial or punishment), and resilience-enhancing ability of each counter-hybrid measure on the basis of an in-depth literature review complemented with a mini-Delphi approach with seven (junior) analysts with a background in strategic studies. This resulted in probability distributions from which samples were drawn to conduct the experiments. Initially, the parameters of these distributions were inspired by a literature review. Subsequently, analysts made a one-time adjustment to the parameters, informed by visualizations of the resulting distributions. The specifics of these probability distributions per variable are summarised in Table \ref{Table1} while the exact parameters are available in the appendix. Values that are likely to be drawn from these probability distributions indicate that they align closely with consensus in the literature and the outcomes of the mini-Delphi survey, while values unlikely to be drawn correspond to values that are less in alignment. By repeatedly sampling input variables from these distributions independently, we generated a total of 1000 experimental scenarios. These experiments can be considered semi-synthetic due to the absence of a rigorous, standardized method for constructing the prior distributions for these estimates \citep{Klami2023}, requiring us to rely on the constructed probabilistic representations.

Despite the semi-synthetic nature of the experiments, all the counter-hybrid measures considered in this paper have been derived from real-world examples and their impacts have been scored by experts, ensuring reflection of real-world variability and available empirical evidence. As the modeling approach enables the exploration of dynamics that cannot be empirically tested in the real world, the semi-synthetic nature of the data is a necessary instrument to conduct this analysis. Furthermore, the flexibility of the proposed framework ensures its applicability to other domains and hybrid threat types, as the underlying principles and interactions are generalizable beyond the specific scenarios tested. This adaptability enhances its utility in addressing a broad spectrum of hybrid threat challenges.

\section{Experimental Design} \label{sectexp}
We consider a scenario in which the defending agent B fears that revisionist agent A attempts to destabilize and harm agent B through hybrid attacks. In particular, the defender is aware of agent A's offensive capabilities in the cyber and information domains and is concerned that the latter will carry out a high-scale cyber-attack against its critical infrastructures, such as power plants, and grids, water management facilities, ports, the healthcare system and/or other essential services. Offensive cyber operations constitute a clear example of a hybrid threat below the threshold of large-scale armed conflict. Indeed, cyber operations have become more prevalent in recent years due to the technical, physical and logical layers of cyberspace and the pervasive use of networks and technologies in our daily life \citep{arce2023cybersecurity}.\footnote{Council of Foreign Relations, “Cyber Operations Tracker,” accessed December 1, 2022, \url{https://www.cfr.org/cyber-operations/} \label{cybertracker}} Furthermore, offensive cyber operations may well produce material consequences resulting in considerable physical damage such as for instance in the case of Stuxnet in Iran (2009), Shamoon in Saudi Arabia (2012) or NotPetya in over sixty countries around the world (2017).

An anonymized list of plausible hybrid actions was constructed based on a series of real-world malicious cyber operations drawn from the updated datasets compiled by Valeriano and Maness~\cite{valeriano2014dynamics}, Stirparo et al.~\cite{stirparo2019apt}, and the Council on Foreign Relations,\textsuperscript{\ref{cybertracker}} as well as on a review of the relevant literature. In addition, given the exponential development of new technologies and the evolving dynamics in current conflicts, this was complemented with expert imagination, in an effort to anticipate potential courses of action (and response), and key variables in the cyber domain were distilled. This resulted in a realistic cyber threat scenario that is specified in the appendix. Plausible counter-hybrid responses are similarly drawn and the experts have selected the top five cross-domain measures to counter malicious cyber attacks, which are summarised in Table \ref{Table2}. 

While it is generally difficult to determine the exact costs and damages resulting from a cyber attack on critical infrastructures \citep{greenberg2019}, experts recognize that a defender's ability to timely detect an attack, and recover from it, significantly affects the overall impact of malicious conduct \citep{MORATO201814, SULLIVAN201730}. Therefore, we consider three categories of impact resulting from cyber attacks, based  on the impact that such attacks may produce on critical infrastructures:
\begin{itemize}
    \item $\theta_1$ Impact from cyber attacks on critical infrastructure leading to a massive disruption of essential services due to the absence of effective detection capabilities and the lack of adequate recovery capabilities.
    \item $\theta_2$ Impact from cyber attacks on critical infrastructure leading to substantial losses due to the defender's ability to partially, or at least eventually, detect the attack or mitigate its consequences.
    \item $\theta_3$ Impact from cyber attacks on critical infrastructure leading to limited or negligible effects by reason of the defender's ability to preventively detect and/or recover from the attack.
\end{itemize}
Since entities targeted by malicious cyber attacks are reluctant to share information about their precise impact, we have constructed heavy-tailed half-normal distributions based on the research of \cite{lis2019cyberattacks} to sample the potential impact, $\theta_1, \theta_2, \theta_3$ for each of the three classes of malicious cyber-attacks. Drawing from distributions with high variance, which are specified in the appendix, helps to account for the variability and lack of consensus on the potential impact of a cyber attack.

\subsection{Counter-Cyber Measures}
When parsing available responses to cyber attacks, the focus is placed on in-domain responses (measures in cyberspace), as well as out-domain (measures in other domains) - such as law enforcement, norm development, public diplomacy, and economic sanctions. Through the lens of cumulative deterrence, some of these counter-hybrid measures intend to mitigate the potential damage of a hostile cyber attack, while others aim to dissuade adversaries from conducting aggressive behaviors by raising their cost-benefit calculation. The five different counter-cyber measures considered are active intelligence sharing $d_1$, boosting cyber resilience at the wider societal level $d_2$, employing offensive cyber capabilities $d_3$, imposing market restrictions $d_4$ and open deterrence messaging through strategic communications $d_5$. These are shown in Table \ref{Table2}. Alternatively, the defender can choose to refrain from engaging in counter-hybrid measures $d_6$. The appendix contains the specifications of the probability distributions that will be used to distill the costs, the probability of successful deterrence and the probability of mitigating damaging effects for each of the different counter-cyber measures.  

\subsubsection{Active Intelligence Sharing}
Active intelligence sharing entails the sharing of intelligence across allies to help detect and attribute attacks. Intelligence sharing yields several benefits, ranging from the promotion and the improvement of situational awareness \citep{WAGNER2019101589} to the development of more refined cyber security strategies \citep{FELEDI2013199}. These benefits contribute to mitigating the damages stemming from cyber attacks. However, there are also costs associated with intelligence sharing, such as the cost of accidentally trusting malicious stakeholders with confidential information \citep{WAGNER2019101589}. When the adversary is not being aware of intelligence sharing, it will not impact its decision to conduct a hybrid operation. 

\subsubsection{Boosting Cyber Resilience at the Wider Societal Level}
As critical infrastructure can rely heavily on private stakeholders, boosting cyber resilience at a wider societal level can decrease the potential impact of a cyber attack on critical infrastructure. This can be achieved through either legislation or public-private partnerships. Partnerships between the public and private sectors are preferred as unilateral legislation might lead to a strong focus on futile compliance efforts \citep{tiirmaa2016building}. The measure enhances cyber resilience and therefore decreases the probability of damaging impacts. Therefore, the measure also contributes to deterring by denial. However, setting up these partnerships can be costly as it draws heavily on scarce cybersecurity experts.

\subsubsection{Employ Offensive Cyber Capabilities}
Offensive cyber attacks are targeted at disrupting, degrading or denying adversaries’ offensive capabilities. The offensive cyber operations can be used to infiltrate the networks to temporarily take the adversary offline and prevent it from using such networks to carry out malicious cyber activities. Alternatively, the measure can serve as a threat to further offensive measures. The ability to execute cyber operations bears costs for facilitating a cyber unit and carrying out an attack. There is an aspect of deterrence by denial and deterrence by punishment. The effectiveness of offensive cyber operation as a deterrence measure is debatable as some scholars argue that in order for the offensive operation to be credible, vulnerabilities in the adversaries' network need to be exploited that, when patched, make the offensive cyber operation less useful \citep{mckenzie2017cyber}.

\subsubsection{Market Restrictions}
In the case of critical infrastructures, a state may decide to ban the use of software, hardware or other ICT products and services produced or supplied by allegedly hostile actors. As a softer version, a state may impose strict due diligence and risk assessment obligations with regard to the procurement of ICT services and products. Market restrictions inevitably bear substantial costs \citep{krolikowski2023non}. However, only allowing trusted parties to run critical infrastructure limits a hostile actor's capacity to take control of the critical infrastructure and therefore enhances resilience. At the same time, the measure can contribute significantly to deterrence by denial efforts as it becomes increasingly hard for the adversary to conduct hybrid operations on systems the hostile actor is not familiar with. In this way, it denies the adversary the ability to carry out attacks.

\subsubsection{Open Deterrence Messaging through Strategic Communication}
A defending state can publish national security doctrinal documents stating responses to particular threats. These responses can be actively communicated through strategic communication channels. From a deterrence-by-punishment perspective, it is clear that the effectiveness of the measure relies heavily on the level of detail and the language used (i.e., the level of retaliatory threats it mentions).

\section{Results} \label{results}

\begin{figure*}[!b]
    \captionsetup{justification=centering}
    \centering
    \includegraphics[width = 5.0in]{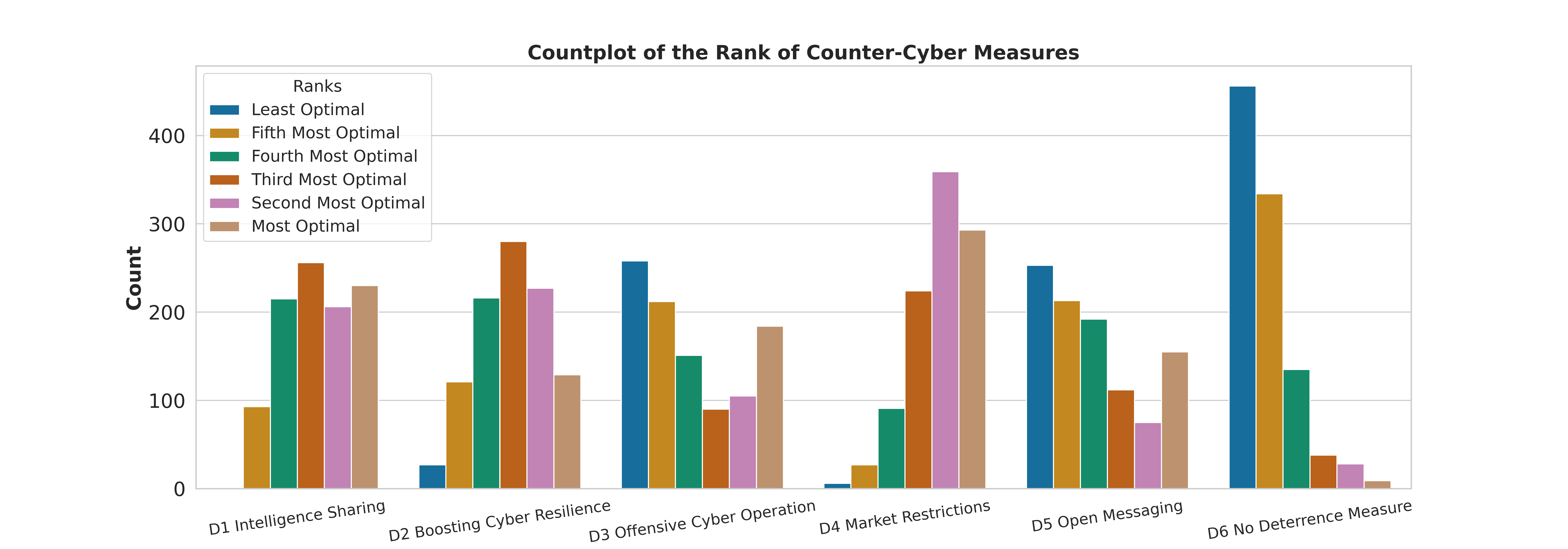}
    \caption{The count of the specific rank that each of the counter-hybrid measures is computed to attain.}
    \label{optmal_sol}
\end{figure*}

In this section, we discuss the results based on the experimental setup of the previous section. First, we present the results of optimizing for the counter-hybrid measure using estimated deterrence probabilities. This is followed by an analysis of the subgame perfect equilibria, where the decision to conduct the hybrid attack is modeled as the agent's strategic choice.\footnote{The modeling effort is publicly available at \url{https://github.com/HCSS-Data-Lab/Hybrid-Threat-Implementation}.}

For each of the 1000 experiments, we rank the effectiveness of the counter-hybrid measure in terms of total pay-off for defending agent B from least optimal to most optimal. A count plot of the rank of the counter-hybrid measures for all experiments is displayed in Figure \ref{optmal_sol}.

In summary, despite the high cost of imposing market restrictions ($d_4$), they are often deemed the most optimal counter-hybrid measure due to the potential to mitigate damages and to deny the adversary's ability to carry out attacks. Intelligence sharing ($d_1$), valued for its cost-effectiveness, plays a crucial role in mitigating attack damage by enabling timely defensive actions and fostering political support for collective responses. While offensive cyber operations ($d_3$) could disrupt enemy capabilities, they carry high risks of escalation and have variable success rates. Boosting cyber resilience ($d_2$), though costly, is consistently rated effective for both damage mitigation and deterrence. Open deterrence communication ($d_5$) hinges on the adversary's responsiveness to threats, requiring detection, attribution, and communication capabilities to be successful. Lastly, in very rare draws, abstaining from counter-hybrid measures ($d_6$) emerges as the most effective strategy.

To derive more meaningful insights, we now shift our focus from the specific outcomes of individual measures to the broader results that can be drawn from the overarching characteristics of these measures, especially considering that a well-designed elicitation protocol would significantly enhance the interest and reliability of individual results while the same overarching characteristics would prevail.

\begin{figure}[b]
    \captionsetup{justification=centering}
    \centering

    \includegraphics[width = 3.5in]{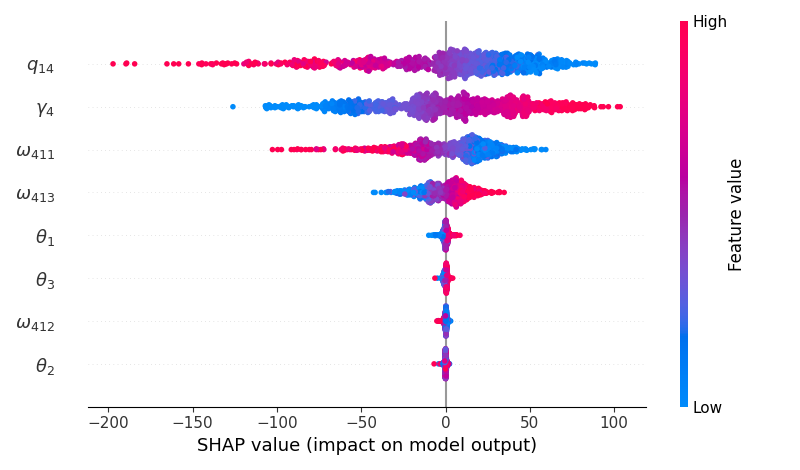}
    \caption{SHAP summary plot for counter-hybrid measure imposing market restrictions: The y-axis represents the features ranked by their importance to the model output. The x-axis shows the SHAP value, indicating the magnitude and direction of each feature's impact on the model output. The color gradient reflects the feature values.}
    \label{shapley}
\end{figure}

Overall, the measures vary in several ways: some measures rely on their ability to dissuade the adversary through punishment (open messaging, offensive cyber operations), others count on the ability to mitigate the potential damage when confronted with an attack (intelligence sharing), and there are also measures that are a mixture of both deterrence by denial and enhancing resilience (boosting cyber resilience and imposing market restrictions). While these characteristics are distributed evenly among the optimal measures, the measure designated as optimal in most cases  - i.e., imposing market restrictions -  is also the most versatile one with respect to both dissuasion as well as resilience enhancement. In addition, the variance of the cost, ability to mitigate the damage and ability to deter are different for each of the measures as their impact is mediated by favoring conditions. For instance, while deterrence by punishment measures (open messaging and offensive cyber operation) can be very effective counter-hybrid measures, they are also among the most ineffective measures for some experiments as illustrated by Figure \ref{optmal_sol}. This is because they rely heavily on their effect on the adversary's strategic calculus. When dissuasion is not successful, they do not contribute to mitigating the damaging impact of hybrid conduct, leaving the defender exposed. 

\begin{table}[!b]
\centering
\caption{Top 3 Most Importance Features of the SHAP Plots for each Counter-Hybrid Measure}
\begin{tabular}{|c|c|c|c|}
\hline
\multirow{2}{*}{Counter-hybrid Measures} & \multicolumn{3}{c|}{Top Features} \\ 
\cline{2-4}
 & \#1 & \#2 & \#3  \\ 
\hline
D1 Intelligence Sharing & $q_{11}$ & $\omega_{111}$ & $\omega_{113}$ \\ 
\hline
D2 Boosting Cyber Resilience & $\gamma_2$ & $q_{12}$ & $\omega_{213}$  \\ 
\hline
D3 Offensive Cyber Operation & $q_{13}$ & $\omega_{311}$ & $\omega_{313}$  \\ 
\hline
D4 Market Restrictions & $q_{14}$ & $\gamma_{4}$ & $\omega_{411}$  \\ 
\hline
D5 Open Messaging & $q_{15}$ & $\gamma_{5}$ & $\theta_{1}$  \\ 
\hline
D6 No Deterrence Measure & $q_{16}$ & $\omega_{611}$ & $\theta_{1}$  \\ 
\hline
\end{tabular}
\label{tab:sensan}
\end{table}

In order to test how variations in input parameters influence the output of the model, we conducted sensitivity analyses using the state-of-the-art tool of van Stein et al. \cite{9903639} for each of the counter-hybrid measures. While full sensitivity reports for each of the counter-hybrid measures are available\footnote{Full sensitivity reports are available at \url{https://github.com/HCSS-Data-Lab/Hybrid-Threat-Implementation}} and the top three contributing features per counter-hybrid measure are summarized in Table \ref{tab:sensan}, Figure \ref{shapley} presents the SHAP summary plot for imposing market restrictions, which serves as a representative example of the broader sensitivity analysis conducted. As can be observed from the figure, the probability of successfully deterring the adversary $q_{14}$, as well as the costs of the measure $\gamma_{4}$ are the most influential factors for the effectiveness of the counter-hybrid measure. While the measure's ability to mitigate the negative impact is comparatively less influential overall, its role in increasing the probability of a negligible impact ($\omega_{413}$) or reducing the likelihood of a severe impact ($\omega_{411}$) remains a significantly important factor in evaluating its effectiveness. As the effectiveness of the measure is strongly influenced by its ability to dissuade the adversary, there is a need to consider this factor not only probabilistically but also within a game-theoretic framework. By doing so, we computed the subgame perfect equilibria detailed in the previous section, providing a more comprehensive evaluation of the measures in a strategic context.

\begin{table}[!t]
\centering
\caption{Subgame Perfect Equilibria Outcome Occurences}
\begin{tabular}{|c|c|c|}
\hline
\multirow{2}{*}{Counter-hybrid Measures} & \multicolumn{2}{c|}{Hybrid Operation} \\ 
\cline{2-3}
 & Attack & No Attack \\ 
\hline
D1 Intelligence Sharing & 15 & 697 \\ 
\hline
D2 Boosting Cyber Resilience & 1 & 98 \\ 
\hline
D3 Offensive Cyber Operation & 3 & 76 \\ 
\hline
D4 Market Restrictions & 0 & 110 \\ 
\hline
D5 Open Messaging & 0 & 0 \\ 
\hline
D6 No Deterrence Measure & 0 & 0 \\ 
\hline
\end{tabular}
\label{tab:spe}
\end{table}

These subgame perfect equilibria for the same 1000 experiments are displayed in Table \ref{tab:spe}, reflecting outcomes where agents seek to optimize their pay-off rationally. The low occurrence of hybrid attacks indicates that the chosen counter-hybrid measures focus on deterring the adversary rather than mitigating the consequences of a hybrid attack. Given that the adversary's strategic calculus is assumed to be known, the defending agent can strategically select the most cost-effective counter-hybrid measure that successfully deters the adversary from launching an attack. This explains why intelligence sharing is preferred over market restrictions when both measures suffice to deter the adversary. Moreover, when the strategic calculus is such that the adversary is likely to proceed with a hybrid operation regardless of the counter-hybrid measure, the subgame perfect equilibria suggest that the defending agent should commit to cost-efficient counter-hybrid measures, such as intelligence sharing, to mitigate the impact of the attack.

\subsection{Validation of the Results}
While general validation of the results remains challenging due to the ambiguous nature of hybrid threats and the reluctance of targeted parties to disclose information, contextualizing our findings within established models and prior studies provides valuable insights. First, our findings align with other model implications that hybrid threats are effective in economically exhausting the defending agent \cite{balcaen2022game}, as the most effective counter-hybrid measure in our analysis is also the most costly. This underscores the need for a prioritized allocation of scarce resources and the strategic construction of cross-domain deterrence measures. The findings of our multi-agent model further corroborate prior research emphasizing the need for defensive strategies that balance the effectiveness of counter-hybrid measures with resource constraints \cite{attiah2018game}, highlighting the importance of selecting cost-efficient options under varying adversarial strategies. 

Moreover, our analysis supports existing studies on democratic deterrence, which advocate for a whole-of-society approach coordinated by the state \cite{treverton2018addressing}. In particular, our findings highlight the effectiveness of measures such as market restrictions and intelligence sharing, which align with the use of non-military, soft power, and asymmetrical strategies to counter hybrid threats \cite{wigell2021democratic}. Finally, our sensitivity analysis reinforces the findings on existing modeling efforts on deterrence in the cyber realm \cite{kostyuk2021deterrence}, emphasizing that the effectiveness of such threats is heavily contingent upon the adversary's susceptibility to countermeasures.

\section{Discussion} \label{discus} 
This paper introduced new approaches to evaluating the effectiveness of cross-domain counter-hybrid measures by balancing the cost, deterrence ability and damage mitigation ability of each of the measures in the context of uncertainty about the deterrent and defensive effects of these measures, which were therefore modeled probabilistically or game-theoretically. Experiments with these approaches included 1,000 different scenarios involving malicious cyber operations and contained various counter-hybrid measures inspired by an extensive literature review complemented with a mini-Delphi survey.

While game-theoretic approaches have gained traction in the hybrid domain \citep{balcaen2022game} and the cyber domain \citep{attiah2018game}, game theory alone fails to capture the subtleties arising from uncertainties inherent in the hybrid threat dynamic. Alternatively, probabilistic graphical models can be inadequate in the hybrid threat domain \citep{balaban2018hybrid}, as they require ex-ante specifications and do not account for strategic interactions. Our work is the first to unify game-theoretic and probabilistic graphical approaches, encoding uncertainties specific to the hybrid domain as probabilistic relations and modeling the different actors as agents. The results of our modeling effort highlight how distinct characteristics of counter-hybrid measures influence their effectiveness across various simulations and illustrate game outcomes shaped by rational, optimal decision-making through game equilibria.

The policy implications are therefore twofold. First, our analysis shows that the effect of different counter-hybrid measures ranging from punishment, denial to resilience can be systematically estimated under conditions of deep uncertainty drawing on insights from the literature and through expert elicitation. The outcome of the multi-agent approach even reveals the changing optimal measure in the case of a strategic-acting adversary. Our model provides a prototype for this process, a model that can be emulated, expanded, and refined to test and design counter-hybrid policies in order to help policy-makers in the formulation and prioritization of counter-hybrid policies.

Second, our extensive simulations grounded in elicited data highlight the importance of favoring conditions. These conditions mediate the effect of measures, suggesting that effective counter-hybrid strategies need to target these favoring conditions. Deterrence by punishment measures, for instance, are contingent on the receptiveness of the adversary, as these counter-hybrid strategies only work well when the adversary is responsive to such measures. \citep{RR-2451-A}. The modeling exercise indicates that even a small enhancement in understanding the aggressor’s plausible receptiveness to counter-hybrid measures could lead to a significant enhancement in the assessment of the effectiveness of measures. This implies that resources spent on anticipating the adversary’s reaction to possible counter-hybrid measures are conditional on the effect of the counter-hybrid measures. Because the responsiveness of the adversary also depends on communication, the results emphasize that a meticulous communication strategy is essential when implementing a counter-hybrid measure \citep{sweijs2021essence}.

We propose two future research areas. First, recently developed methods have focussed on the elicitation of conditional probability tables \citep{HASSALL2019104539,barons2022balancing}. Implementation of these methods and confronting policy-makers would increase the precision of the input data for the model. Second, experiments should be run on scenarios that involve hybrid threats of a cross-domain nature beyond just the cyber domain. This injects additional complexity but will reflect more closely the contemporary nature of hybrid threats. 

More fundamentally, our modeling effort seeks to develop the knowledge base including the methods, data, and techniques, that can be applied to real-world security problems characterized by uncertainty. By showing that such a modeling effort yields insights that have real-world policy value, we hope our effort will receive a wider following. 

\section*{Funding}
This work was financially supported by the Dutch Ministry of Foreign Affairs and the Dutch Ministry of Defence within the PROGRESS research framework agreement. Responsibility for the contents and for the opinions expressed rests solely with the authors and does not constitute, nor should be construed as, an endorsement by the Dutch Government.

\section*{Disclosure Statement}
No potential conflict of interest was reported by the authors.

\bibliography{sn-bibliography.bib}
\clearpage

\section{Appendix}
\section{Cyber threat scenario} \label{scenariocyber}

Amidst times of intense crisis in which political and economic relations are currently being challenged, state B fears that revisionist states, such as state A, may attempt to destabilize and influence B with a combination of hybrid threats and attacks. In particular, state B is concerned that - given the established power of state A in the information and cyber domains - the latter would carry out a high-scale cyber attack against its critical infrastructures, including power grids and ports.  

Intelligence services of state B reported that B is likely to be the next target of a cyber attack on critical infrastructures and flagged the latter circumstance as a credible threat. 

While several countries similar to state B have already been the target of malicious cyber operations by state A, some of them have been successful in deterring or mitigating the impact of these attacks by means of cross-domain approaches, either by setting red lines and clearly communicating countermeasures (deterrence by punishment) or by enhancing resilience and raising the costs of attacks to dissuade the adversary (deterrence by denial).  

Despite the deterrence efforts taken by target countries, state A still decided to conduct an offensive cyber operation against some of the countries. Now, the potential impact of these cyber operations for each of the targeted countries differs heavily, also depending on the detection capabilities of the targeted countries, as well as on their prevention and recovery capabilities. Where detection capabilities were mediocre and failed to preventively identify the threat, vulnerabilities were exploited more profoundly by the adversary, eventually causing a significant harmful impact.  

Adversary A decided to conduct a cyber operation based on the potential gains and losses. Surely, A takes into account that the hybrid operation is more successful when not properly detected. Therefore, A carefully examines the detection and attribution ability of defender B given the envisioned cyber attack. Defender B, in turn, tries to enhance its detection and attribution ability when enhancing its resilience.  

\section{Experimental Design Data} \label{expdes}
This section specifies the details about the experiments of the experimental section by specifying the probability distributions these experiments are drawn from.

We draw the most damaging impact $\theta_1$ and the substantial impact $\theta_2$ from the truncated normal distributions $f(x,\mu_1,\sigma_1,a,b)$ and $f(x,\mu_2,\sigma_2,a,b)$ respectively, where $a=0$ , $\mu_1,\mu_2=1000,100$ and $\sigma_1,\sigma_2=300, 50$ respectively \cite{norgaard1980expected}. Hence $x$ is drawn from the interval $[0,\infty]$. Negligible damaging impact $\theta_3$ is drawn from a positive half-normal distribution based on normal distribution  $\mathcal{N}(0,5)$. All values represent millions of damaging costs.

The costs $\gamma_i$ for each of the counter-hybrid measures $d_i$ are drawn from truncated normal distributions. The probability of the adversary conducting a hybrid operation $q_{ij}$ after counter-hybrid measure $d_i$ is drawn from beta distributions as it is the conjugate prior to the Bernoulli distribution (assuming the adversary either attacks or does not attack). Finally, the probabilities of the potential impact of hybrid conduct $w_{ijh}$ based on counter-hybrid measure $d_i$ and attack $c_j$ are drawn from a Dirichlet distribution as it is the conjugate prior to the categorical distribution and commonly used in influence diagrams \citep{spiegelhalter1990sequential}. 

\subsection{Active intelligence sharing}
When intelligence is shared, this is not directly communicated to the adversary, leaving the probability that the adversary conducts a hybrid operation after this measure $q_{11}$ at $Be(5,5)$. However, this measure significantly improves the mitigation ability of the damaging impact of the hybrid conduct leaving the probabilities being drawn from Dirichlet distribution:
\begin{align*}
\left(
\begin{matrix}
w_{111} \\ w_{112} \\ w_{113} 
\end{matrix} 
\right)
\sim\text{Dir}
\left(
\begin{matrix}
4 \\ 8 \\ 12  
\end{matrix}\right).
%\quad \quad  \quad 
\end{align*}
The cost of this measure accounts for losing confidential information to not entitled parties and is drawn from the truncated normal distributions $f(x,\mu,\sigma,a,b)$ with $\mu=150, \sigma=50$ and $a=0$.

In case the adversary does not conduct a hybrid operation, the probability that the damaging impact will be negligible is always 1.

\subsection{Boost cyber resilience at the wider societal level}
Boosting cyber resilience works via deterrence by denial leaving the probability that the measure dissuades the adversary from committing a hybrid conduct $q_{21}$ drawn from $Be(4,8)$ and the mitigation of potential impact drawn from the following Dirichlet distributions:
\begin{align*}
\left(
\begin{matrix}
w_{211} \\ w_{212} \\ w_{213} 
\end{matrix} 
\right)
\sim\text{Dir}
\left(
\begin{matrix}
3 \\ 5 \\ 8  
\end{matrix}\right).
%\quad \quad  \quad 
\end{align*}
The cost of this measure is mostly carried by the private sector and is drawn from truncated normal distribution $f(x,\mu,\sigma,a,b)$ with $\mu=300, \sigma=50$ and $a=0$.

\subsection{Offensive Cyber Operation}
An offensive cyber operation has deterrence by denial as well as deterrence by punishment components. To compensate for the fact that the measure can also backfire, we draw the probability that this measure is successful in dissuading the adversary from committing hybrid conduct $q_{31}$ from $Be(1,1.2)$ and the damage mitigation potential drawn from the following Dirichlet distributions:
\begin{align*}
\left(
\begin{matrix}
w_{311} \\ w_{312} \\ w_{313} 
\end{matrix} 
\right)
\sim\text{Dir}
\left(
\begin{matrix}
1 \\ 1 \\ 1 
\end{matrix}\right).
%\quad \quad  \quad 
\end{align*}
The costs of this measure contain the cost of setting up an offensive cyber unit as well as the cost that comes with the associated attack. It is drawn from truncated normal distribution $f(x,\mu,\sigma,a,b)$ with $\mu=250, \sigma=30$ and $a=0$.

\subsection{Market restriction}
Imposing market restrictions works via deterrence by denial and therefore the probability that this counter-hybrid measure dissuades the adversary from committing hybrid conduct $q_{41}$ is drawn from $Be(2,8)$. The probabilities of potential impacts are drawn from the Dirichlet distribution
\begin{align*}
\left(
\begin{matrix}
w_{411} \\ w_{412} \\ w_{413} 
\end{matrix} 
\right)
\sim\text{Dir}
\left(
\begin{matrix}
2 \\ 2 \\ 15 
\end{matrix}\right).
%\quad \quad  \quad 
\end{align*}
Finally, the costs of the measure involve excluding certain private organizations from the market and is drawn from truncated normal distribution $f(x,\mu,\sigma,a,b)$ with $\mu=400, \sigma=50$ and $a=0$.

\subsection{Open deterrence messaging through strategic communications}
Assuming that the message involves some deterrence by punishment and there is uncertainty involved in threatening, we draw the probability that this deterrence measure successfully dissuades the adversary from committing a hybrid conduct $q_{51}$ from $Be(0.4,2)$. 
Damage mitigation is not involved and therefore the probabilities for damaging impacts are drawn from the same Dirichlet distribution as no measure was taken:
\begin{align*}
\left(
\begin{matrix}
w_{511} \\ w_{512} \\ w_{513} 
\end{matrix} 
\right)
\sim\text{Dir}
\left(
\begin{matrix}
12 \\ 6 \\ 2 
\end{matrix}\right).
%\quad \quad  \quad 
\end{align*}
Finally, the costs of the measure also include costs of a risky escalation that one commits to and is drawn from truncated normal distribution $f(x,\mu,\sigma,a,b)$ with $\mu=500, \sigma=250$ and $a=0$.

\subsection{No deterrence measure}
Assuming no deterrence measure taken $d_6$, we draw the probability that the adversary conducts a hybrid operation $q_{61}$ from $Be(5,5)$. Similarly, in case the defender does not conduct a deterrence effort and the adversary conducts a hybrid operation the probability of each of the three impacts is drawn from the Dirichlet distribution:

\begin{align*}
\left(
\begin{matrix}
w_{611} \\ w_{612} \\ w_{613} 
\end{matrix} 
\right)
\sim\text{Dir}
\left(
\begin{matrix}
12 \\ 6 \\ 2 
\end{matrix}\right).
%\quad \quad  \quad 
\end{align*}
In case the adversary does not conduct a hybrid operation, the probability that the damaging impact will be negligible is always 1.

\begin{figure}[]
    \centering
    \includegraphics[width = 3.2in]{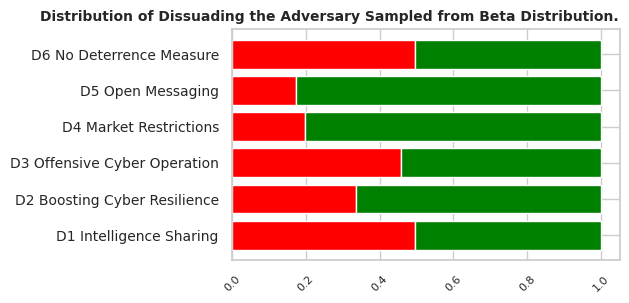}
    \caption{The probability that each of the counter-hybrid measures succeeds in deterring the adversary. While the green indicates the probability that the adversary is successfully dissuaded, the red illustrates the probability that the adversary still conducts a cyber operation.}
    \label{adversaryprobs}
\end{figure}

\begin{figure}[]
    \centering  
    \includegraphics[width = 3.2in]{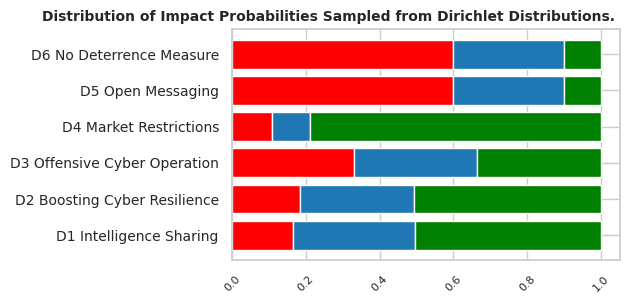}
    \caption{The probability that each of the counter-hybrid measures succeeds in mitigating the hybrid operation. While the red indicates the probability the hybrid operation has a severe impact, the blue indicates the operation has a mediocre impact and the green indicates the hybrid operation has negligible impact.}
    \label{mitigatingprobs}
\end{figure}

Figure \ref{adversaryprobs} and \ref{mitigatingprobs} illustrate the distribution of successful deterrence and the distribution of the potential impact of the malicious cyber attack, respectively.

\end{document}